\documentclass{article}
\usepackage{amsmath}
\usepackage{amsfonts}
\usepackage{latexsym}
\usepackage{graphicx}
\usepackage{url}
\usepackage{colortbl}
\usepackage{rotating} 
\usepackage{setspace}
\DeclareFontFamily{OT1}{pzc}{}
\DeclareFontShape{OT1}{pzc}{m}{it} {<-> s * [1.2] pzcmi7t}{}
\DeclareMathAlphabet{\mathpzc}{OT1}{pzc}{m}{it}
\newcommand{\db}{\mathpzc{d}_\mathcal{B}}
\newcommand{\R}{\mathbb{R}}

\begin{document}
\title{FAST REDSHIFT CLUSTERING WITH THE BAIRE (ULTRA) METRIC}
\author{F. Murtagh $^{*1}$ and P. Contreras$^{*2}$ \\
* Department of Computer Science, Royal Holloway University of London,\\
Egham, Surrey TW20 0EX, UK\\
Email: fmurtagh@acm.org, pedro@cs.rhul.ac.uk \\
1 Science Foundation Ireland, Wilton Park House, Wilton Place\\
Dublin 2, Ireland \\
2 Thinking SAFE Ltd., Egham, Surrey TW20 0EX, UK}

\date{}
\maketitle

\begin{abstract}
The Baire metric induces an ultrametric on a dataset and is of linear computational complexity, contrasted with the standard quadratic time agglomerative hierarchical clustering algorithm.  
We apply the Baire distance to spectrometric and photometric redshifts from the  Sloan Digital Sky Survey using, in this work, about half a million astronomical objects.   We want to know how well the (more costly to determine) spectrometric redshifts can predict the (more easily obtained) photometric redshifts, i.e.\ we seek to regress the spectrometric on the photometric redshifts, and we 
develop a clusterwise nearest neighbor regression procedure for this.   
\end{abstract}



\section{Introduction: Agglomerative Hierarchical Clustering Algorithms}

A metric space $(X,d)$ consists of a set~$X$ on which is defined a 
distance function $d$ which assigns to each pair of points of 
$X$ a distance between them, and satisfies the following four axioms 
for any triplet of points $x, y , z$:

$\mbox{\textbf{A1:} } \forall x, y \in X,  d(x,y) \geq 0 \mbox{ (positiveness)}$\smallskip

$\mbox{\textbf{A2:} } \forall x, y \in X, d(x,y) = 0 \mbox{ iff } x = y \mbox{ (reflexivity)}$\smallskip

$\mbox{\textbf{A3:} } \forall x, y \in X, d(x,y) = d(y,x)  \mbox{ (symmetry)}$\smallskip

$\mbox{\textbf{A4:} } \forall x, y, z \in X, d(x,z) \leq d(x,y) + d(y,z)  \mbox{ (triangle inequality)}$\\

When considering an ultrametric space we need to consider the strong triangular inequality or ultrametric inequality defined as: \smallskip

$\mbox{\textbf{A5:} } d(x,z) \leq max~\{d(x,y), \ d(y,z)\} \mbox{ (ultrametric inequality)}$\smallskip

\noindent
and this in addition to the positivity, reflexivity and symmetry properties (properties A1, A2, A3) 
for any triple of point  $x, y, z \in X$.  

If $X$ is endowed with a metric, then this metric can be mapped onto an ultrametric.  In practice, endowing  $X$ with a metric can be relaxed to a dissimilarity.  An often used mapping from metric to ultrametric is by means of an agglomerative hierarchical clustering algorithm.  A succession of $n - 1$ pairwise merge steps take place by making use of the closest pair of singletons and/or clusters at each step.  Here $n$ is the number of observations, i.e.\ the cardinality of set $X$.  
Closeness between singletons is furnished by whatever distance or dissimilarity is in use. For closeness between singleton or non-singleton clusters, we need to define an inter-cluster distance or dissimilarity.  This can be defined with reference to the cluster compactness or other property that we wish to optimize at each step of the algorithm.   

Since agglomerative hierarchical clustering requires consideration of pairwise dissimilarities at each stage it can be shown that even in the case of the most efficient algorithms, e.g.\ those based on reciprocal nearest neighbors and nearest neighbor chains~\cite{Murtagh85-1}, $O(n^2)$ or quadratic computational time is required.  The innovation in the work we present here is that we carry out hierarchical clustering in a different way such that $O(n)$ or linear computational time is needed.  As always in computational theory, these are worst case times.  

A hierarchy, $H$, is defined as a binary, rooted, node-ranked tree, also termed a dendrogram \cite{benz,john,Lerman81,Murtagh85-1}.
A hierarchy defines a set of embedded subsets of a given set of objects $X$, indexed by the set $I$. These subsets are totally ordered by an index function $\nu$, which is a stronger condition than the partial order required by the subset relation. A bijection exists between a hierarchy and an ultrametric space.

Let us show these equivalences between embedded subsets, hierarchy, and binary tree, through the constructive approach of inducing $H$ on a set $I$.

Hierarchical agglomeration on $n$ observation vectors with indices $i \in I$ involves a series of $1, 2, \dots , n-1$ pairwise agglomerations of observations or clusters, with the following properties.  A hierarchy $H = \{ q | q \in 2^I \} $ such that (i) $I \in H$, (ii) $i \in H \ \forall i$, and (iii) for each $q \in H, q^\prime \in H: q \cap q^\prime \neq \emptyset \Longrightarrow q \subset  q^\prime \mbox{ or }  q^\prime  \subset q$.  Here we have denoted the power set of set $I$ by $2^I$. An indexed hierarchy is the pair $(H, \nu)$ where the positive function defined on $H$, i.e., $\nu : H \rightarrow \R^+$, satisfies: $\nu(i) = 0$ if $i \in H$ is a singleton; and (ii)  $q \subset  q^\prime \Longrightarrow \nu(q) < \nu(q^\prime)$.  Here we have denoted the positive reals, including 0, by $\R^+$. Function $\nu$ is the agglomeration level.  Take  $q \subset  q^\prime$, let $q \subset q''$  and $q^\prime \subset q''$, and let $q''$ be the lowest level cluster for which this is true. Then if we define $D(q, q^\prime) = \nu(q'')$, $D$ is an ultrametric.  

In practice, we start with a Euclidean or alternative dissimilarity, use some criterion such as minimizing the change in variance resulting from the agglomerations, and then define $\nu(q)$ as the dissimilarity associated with the  agglomeration carried out.

\section{Baire or Longest Common Prefix Distance}
\label{section:baire}

Agglomerative hierarchical clustering algorithms are constructive hierarchy-constructing algorithms.  Such algorithms have the aim  of mapping data into an ultrametric space, or searching for an ultrametric embedding, or ultrametrization~\cite{Rooij78}. 

Now, inherent ultrametricity leads to an identical result with most commonly used agglomerative criteria~\cite{Murtagh85-1}. Furthermore, data coding can help greatly finding how inherently ultrametric data is~\cite{Murtagh04}.  In certain respects the hierarchy determined by the Baire distance can be viewed as a particular coding of the data because it seeks longest common prefixes in pairs of (possibly numerical) strings.  
We could claim that determining the longest common prefix is a form of data compression because we can partially express one string in terms of another.  

\subsection{Ultrametric Baire Space}
\label{subsection:baire}

A Baire space consists of countably infinite sequences with a metric defined in terms of the longest common prefix: the longer the common prefix, the closer a pair of sequences. What is of interest to us here is this longest common prefix metric, which we call the Baire distance~\cite{Murtagh08, Contreras07}. 

Consider real-valued or floating point data (expressed as a string of digits rather than some other form, e.g.\ using exponent notation).   
The longest common prefixes at issue are those of precision of any value.  For example, let us consider two such values, $x_{i}$ and $y_{j}$, with $i$ and $j$ ranging over numeric digits.   When the context easily allows it, we will call these $x$ and $y$. 

We take $x$ and $y$ to be bounded by 0 and 1. Each are of some precision, and we take the integer $\left| K \right|$ to be the maximum precision.   

Thus we consider ordered sets $x_{k}$ and $y_{k}$ for $k \in K$. In line with our notation, we can write $x_{k}$ and $y_{k}$ for these numbers, with the set $K$ now ordered. So, $k = 1$ is the first decimal place of precision; $k = 2$ is the second decimal place; . . . ; $k = \left| K \right|$ is the $\left| K \right|th$ decimal place.  The cardinality of the set K is the precision with which a number, $x_{k}$, is measured. 

Take as examples $x_{k} = 0.478$; and $y_{k} = 0.472$. In these cases, $\left| K \right| = 3$. Start from the first decimal position.
For $k = 1$, we find $x_{k} = y_{k} = 4$. For $k = 2$, $x_{k} = y_{k}$ . But for $k = 3$,  $x_{k} \neq y_{k}$.

We now introduce the following distance (case of vectors $x$ and $y$, with 1 attribute, hence unidimensional):

\begin{equation}
\label{eq:baire}
\db(x_{K}, y_{K}) =   
	\left\{ 
	\begin{array}{ll}
       1 &\;\; $if$\;\;  x_{1} \neq y_{1}\\
       $inf$\;\;  2^{-k} & \;\;\;\;\;\; x_{k} = y_{k} 
                        \;\;\; 1 \leq k \leq
       \left| K \right|
    \end{array}
    \right.
\end{equation}
We call this $\db$ value Baire distance, which is an ultrametric~\cite{Murtagh04, Murtagh04-2, Murtagh04-1, Murtagh05, Murtagh08} distance.

Note that the base 2 is given for convenience. When dealing with binary data $x, y$, then 2 is the chosen base. When working with real numbers the base can be redefined to 10.

\subsection{Agglomerative Hierarchical Clustering Algorithm versus Hierarchical Encoding of Data}

The Baire distance was introduced and described by Bradley~\cite{bradley} in the context of inducing a hierarchy on strings over finite alphabets.  This work further pursued the goal of embedding a dendrogram in a p-adic Bruhat-Tits tree, informally characterized as a ``universal  dendrogram''.  

By convention we denote a prime by p, and a more general, prime or non-prime, positive integer by m.  

A geometric foundation for ultrametric structures is presented in  Bradley \cite{bradley0}.   Starting from the point of view that a  dendrogram, or ranked or unranked, binary or more general $m$-way, tree, is an object in a p-adic geometry, it is noted that:  ``The consequence of using p-adic methods is the shift of focus from imposing a hierarchic structure on data to finding a p-adic encoding which reveals the inherent hierarchies.''

This summarizes well our aim in this work.  We seek hierarchy and rather than using an agglomerative hierarchical clustering algorithm which is of quadratic computational time (i.e., for $n$ individuals or observation vectors, $O(n^2)$ computational time is required) we instead seek to read off a p-adic or m-adic tree.  (We recall that $n$ is the number of observations, or cardinality of the observation set $X$.) 
In terms of a tree, p-adic or m-adic mean p-way or m-way, respectively, or that each node in the tree has at most p or m, respectively, sub-nodes.  

Furthermore, by ``reading off'' we are targeting a linear time, or $O(n)$ algorithm involving one scan over the dataset, and we are imposing thereby an encoding of the data.  

In practice we will be more interested in this work in the hierarchy, and the encoding algorithm used is a means towards this end.  For a focus on the encoding task, see Murtagh~\cite{steklov}.

\subsection{A Baire-Based Hierarchical Clustering Algorithm}
\label{subsection:alg}

A straightforward algorithm for hierarchical clustering based on the Baire distance, as described in section \ref{subsection:baire} is as follows. 
Because of working with real numbers in our case study below, we define the base in relation \ref{eq:baire} as 10 rather than 2.  

For the first digit of precision, $k = 1$, consider 10 ``bins'' corresponding to the digits $0, 1, \dots , 9$.  For each of the nodes corresponding to these bins, consider 10 subnode bins corresponding to the second digit of precision,  $k = 2$, associated with $0, 1, \dots , 9$ at this second level.  We can continue for a third and further levels.  In practice we will neither permit not wish for a very deep (i.e., with many levels) storage tree.  For the base 10 case, level one (corresponding to $k = 1$) gives rise to up to 10 clusters.   For level two (corresponding to $k = 2$) we have up to 100 clusters.  We see that in practice a small number of levels will suffice.  
In one pass over the data we map each observation (recall that it is univariate but we are using its ordered set of digits, i.e.\ ordered set $K$) to its bin or cluster at each level.   For $\ell$ levels, the computation required is $n \cdot \ell$ operations.   For a given value of $\ell$ we therefore have $O(n)$ computation -- and furthermore with a very small constant of proportionality since we are just reading off the relevant digit and, by design, updating a node or cluster membership list and cardinality.  

\section{Inducing a Hierarchy on the SDSS Data using the Baire Ultrametric}
The Sloan Digital Sky Survey (SDSS)~\cite{SDSS} is systematically mapping the sky producing a detailed image of it and determining the positions and absolute brightnesses of more than 100 million celestial objects. It is also measuring the distance to a million of the nearest galaxies and to a hundred thousand quasars.  See Adelman-McCarthy et al.~\cite{Adelman-McCarthy07} for a description of the data available in this catalog.

The aim here is to build a mapping from $z_{spec} \rightarrow z_{phot}$ to help calibrating the redshifts, based on the $z_{spec}$ observed values. Traditionally we could map $ f : z_{phot} \longrightarrow z_{spec}$  based on trained data. The mapping $f$ could be linear (e.g.\ linear regression) or non-linear (e.g.\ multilayer perceptron) as used by D'Abrusco~\cite{Dabrusco06-1}. These techniques are  global. Here our interest is to develop a locally adaptive approach based on numerical precision.  That is the direct benefit of the (very fast, hierarchical) 
clustering based on the Baire distance.

We use four parameters: right ascension (RA), declination (DEC), spectrometric ($z_{spec}$) and photometric ($z_{phot}$) redshift. The spectrometric technique uses the spectrum of electromagnetic radiation (including visible light) which radiates from stars and other celestial objects.  The photometric technique uses a faster and economical way of measuring the redshifts.

\subsection{Clustering SDSS Data}
\label{subsection:clustering-based-baire}

In order to perform the clustering process introduced in section~\ref{subsection:baire} and further described in \ref{subsection:alg}, we compare every $z_{spec}$ and $z_{phot}$ data point searching for common prefixes based on the longest common prefix (see section~\ref{subsection:baire}). Thereafter, the data points that have digit coincidences are grouped together to form clusters.

\begin{figure}[hb!]
    \includegraphics[scale=.6]{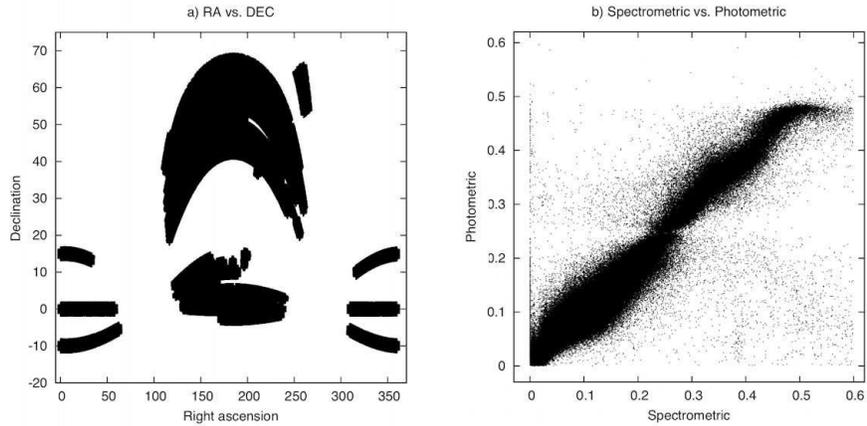}
    \caption{Left: right ascension (RA) versus declination (DEC); Right: $z_{spec}$ versus $z_{phot}$. SDSS data selection used for redshift analysis.}
  \label{fig:characterization}
\end{figure}

Data characterization is presented in Figure~\ref{fig:characterization}. The left panel shows the $z_{spec}$ and $z_{phot}$ sky coordinates of the data currently used by us to cluster redshifts. This section of the sky presents approximately 0.5 million object coordinate points. As can be observed, various sections of the sky are represented in the data. We find this useful since preliminary data exploration has shown that correlation between $z_{spec}$ and $z_{phot}$ is consistent in different parts of the sky. 

This leads us to conclude  that digit coincidences of the redshift measures are distributed approximately uniformly in the sky and are not concentrated spatially. The same occurs for all the other clusters.
We will concentrate on the very near astronomical objects, represented by redshifts between $0$ and $0.6$.  When we plot $z_{spec}$ versus $z_{phot}$ we obtain a highly correlated signal as shown in Figure~\ref{fig:characterization}, right panel. The number of observations that we therefore analyze is 443,014.

\begin{figure}[ht!]
  \begin{center}
    \includegraphics[scale=.6]{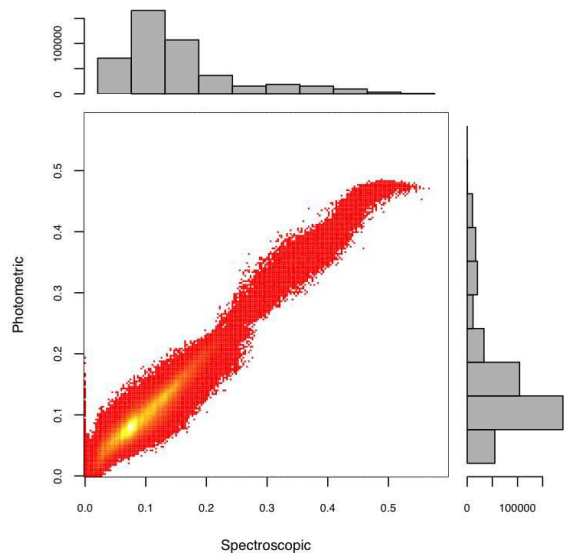}
    \caption{Heat plot and histogram for $z_{spec}$ versus $z_{phot}$. Histogram at the top shows the $z_{spec}$ frequencies, histogram at the right shows $z_{phot}$ frequencies.}
  \label{fig:heat}
  \end{center}
\end{figure}	

Looking at Figure~\ref{fig:heat} it can be seen clearly that most data points fall in the range between 0 and 0.2. Here the histogram on the top shows the $z_{spec}$ data points distribution, and the histogram on the right the $z_{phot}$ data points distribution.  The heat plot (left panel) also highlights the area where data points are concentrated, where the yellow color (white region in monochrome print) shows the major density. 

Consequently, now we know that most cluster data points will fall within this range (0 and 0.2) if common prefixes of digits in the redshift values, taken as strings, are found.   

\begin{figure}[h!]
\centering
	\includegraphics[scale=.6]{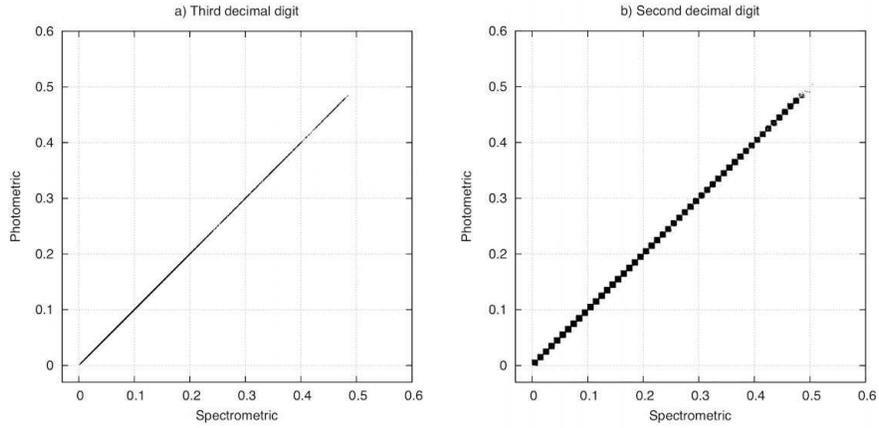}
    \caption{Prefix-wise clustering frequencies depicting 3rd decimal digit coincidences (left panel), and two decimal digit coincidences (right panel).}
	\label{fig:baire-4-3}
\end{figure}

\begin{figure}[h!]
\centering   
	\includegraphics[scale=.6]{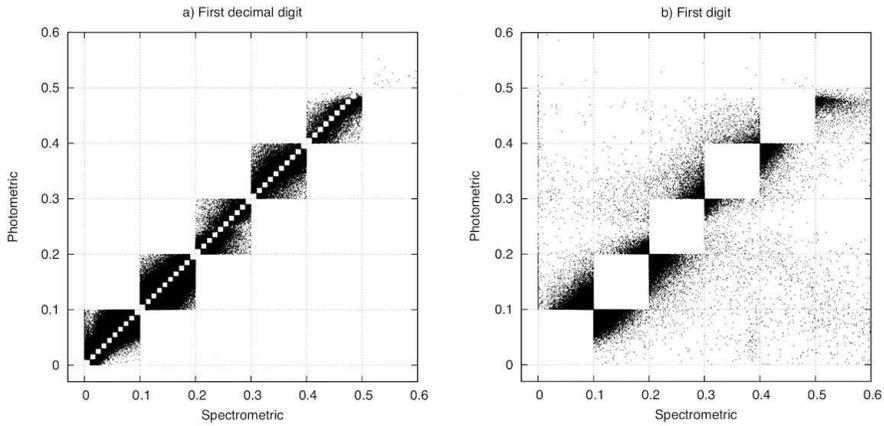}
    \caption{Prefix-wise clustering frequencies depicting the 1st decimal digit coincidences (left panel), and first digit coincidences (right panel).}
  \label{fig:baire-2-1}  
\end{figure}

Figures~\ref{fig:baire-4-3} and~\ref{fig:baire-2-1} show graphically how $z_{spec}$ and $z_{phot}$ correspondences look at different levels of decimal precision. On one hand we find that values of $z_{spec}$ and $z_{phot}$ that have equal precision up to the 3rd decimal digit are highly correlated. On the other hand when $z_{spec}$ and $z_{phot}$ have only the first digit in common, correlation is weak.
For example, let us consider the following situations for plots~\ref{fig:baire-4-3} and~\ref{fig:baire-2-1}:

\begin{itemize}
	\item Figure~\ref{fig:baire-4-3} left: let us take the values of  $z_{spec} = 0.437$ and $z_{phot}= 0.437$. We have that they share the first digit, the first decimal digit, the second decimal digit, and the third decimal digit. Thus, we have a highly correlated signal of the data points that share only up to the third decimal digit.
	\item Figure~\ref{fig:baire-4-3} right: let us take the values of  $z_{spec} = 0.437$ and $z_{phot}= 0.439$. We have that they share the first digit, the first decimal digit, and the second decimal digit. Therefore, the plot shows data points that share only up to the second decimal digit.
	\item Figure~\ref{fig:baire-2-1} left: let us take the values of  $z_{spec} = 0.437$ and $z_{phot}= 0.474$. We have that they share the first digit, and the first decimal digit. Thus, the plot shows data points that share only up to the first decimal digit.
	\item Figure~\ref{fig:baire-2-1} right: let us take the values of  $z_{spec} = 0.437$ and $z_{phot}= 0.571$. We have that they share only the integer part of the value, and that alone. Furthermore, this implies redshifts that do not match in succession of decimal digits.	 For example, if we take the values 0.437 and 0.577, the fact that the third digit is 7 in each case is not of use. 
\end{itemize}

\begin{figure}[hb!]
  \begin{center}
    \includegraphics[scale=.51]{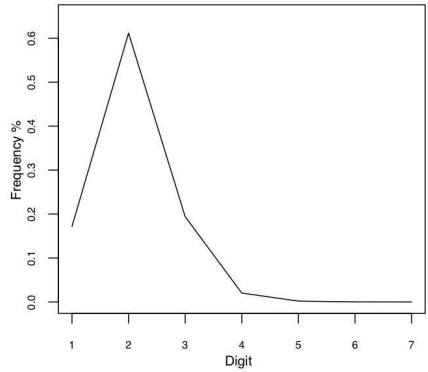}
    \caption{Frequency distribution for Table~\ref{tab:prefixwiseClustering}. The abscissa shows the digit positions, where 1 is the first digit, 2 the first decimal digit, 3 the second decimal digit and so on.}
  \label{fig:freq}
  \end{center}
\end{figure}

Table~\ref{tab:prefixwiseClustering} (see also Figure \ref{fig:freq}) 
shows the clusters found for all different levels of precision. In other words this table allows us to define empirically the confidence levels for mapping of  $z_{phot}$ and $z_{spec}$. For example, we can expect that $82.8\%$ of values for $z_{spec}$ and $z_{phot}$ have at least two common prefix digits. This percentage of confidence is derived as follows: the data points that share six, five, four, three, two, and one decimal digit (i.e., $4 + 90 + 912 + 8,982 + 85,999 + 270,920 = 366,907$ data points. Therefore 82.8\% of the data). 
Additionally we observe that around a fifth of the observations share at least 3 digits in common. Namely, $4 + 90 + 912 + 8,982 + 85,999  = 95,987$ data points, which equals $21.7\%$ of the data.
In previous work~\cite{contreras09} we have shown that the clusters generated by means of the Baire ultrametric are similar to the clusters generated by $k$-means, also see Contreras~\cite{pedrophd}.
\begin{table}[ht!]
\centering
\scalebox{0.8}{
\begin{tabular}{|c|r|c|}
	\hline Digit          & No. \hspace{5pt}& \%     \\\hline
	\hline   1	          &  76,187	        & 17.19  \\\hline 
	\hline  Decimal digit & No. \hspace{5pt}& \%     \\\hline
	\hline   1	          & 270,920      	& 61.14  \\\hline 
	         2	          &  85,999	        & 19.40  \\\hline 
	         3	          &   8,982         	&  2.07  \\\hline 
	         4	          &     912      	&  0.20  \\\hline
	         5	          &      90      	&  0.02  \\\hline 
	         6	          &       4	        &  ---   \\\hline
	  \hline              &  443,094        & 100    \\\hline 
\end{tabular}} 
\caption{Data points based on the longest common prefix for different levels of precision.} 
\label{tab:prefixwiseClustering}
\end{table}
\clearpage


\section{Spectrometric and Photometric Digit Distribution}\label{section:digit-distribution}

We have seen that the Baire ultrametric produces a strict hierarchical classification. In the case of $z_{spec}$ and $z_{phot}$ this can be seen as follows. Let us take any observed measurement of either case of $z_{spec} = z_{phot}$. Let us say $z_{spec} = z_{phot} = 0.1257$. Here we have that for $|K|=4$,  $z_{spec} = z_{phot}$. Hierarchically speaking we have that the root node is 0, for the first level where there potentially exist 6 nodes (i.e. 0,1,...,5); for the second level potentially there are 60 nodes; and so on until $k = |K| = 4$, and $z_{spec} = z_{phot}$, where potentially there are $6\cdot 10 \cdot 10 \cdot 10 = 6000$ nodes. 

Of course not all nodes will be populated.  In fact we can expect that a large number of these potential nodes will be empty if the number of observations $n$ is lower than the potential number of nodes for a certain precision $|K|$ (i.e.\
 $n \leq |K|^{10}$).  Note that this points to a big storage cost,  but in practice the tree is very sparsely populated and $|K|$ small.

A particular interpretation can be given in the case of an observed data point. Following up the above example if we take  $z_{spec} = z_{phot} = 0.1257$, a tree can be produced to store all observed data that falls within this node. Doing this has many advantages from the viewpoint of storing. Access and retrieval, for example, are very fast and it is easy to retrieve all the observations that fall within a given node and its children.

\begin{figure}[h!]
  \begin{center}
    \includegraphics[scale=.8]{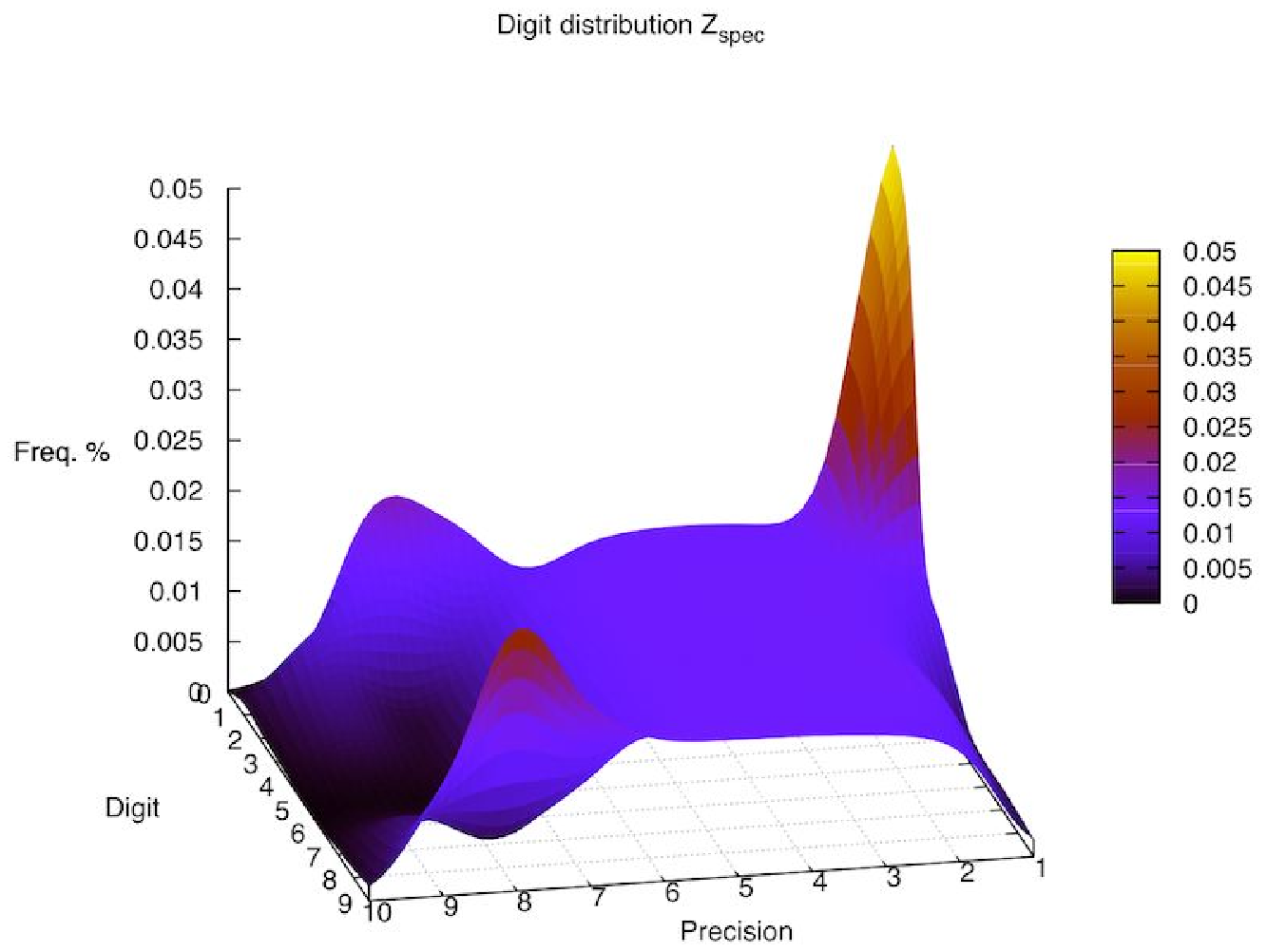}
    \includegraphics[scale=.8]{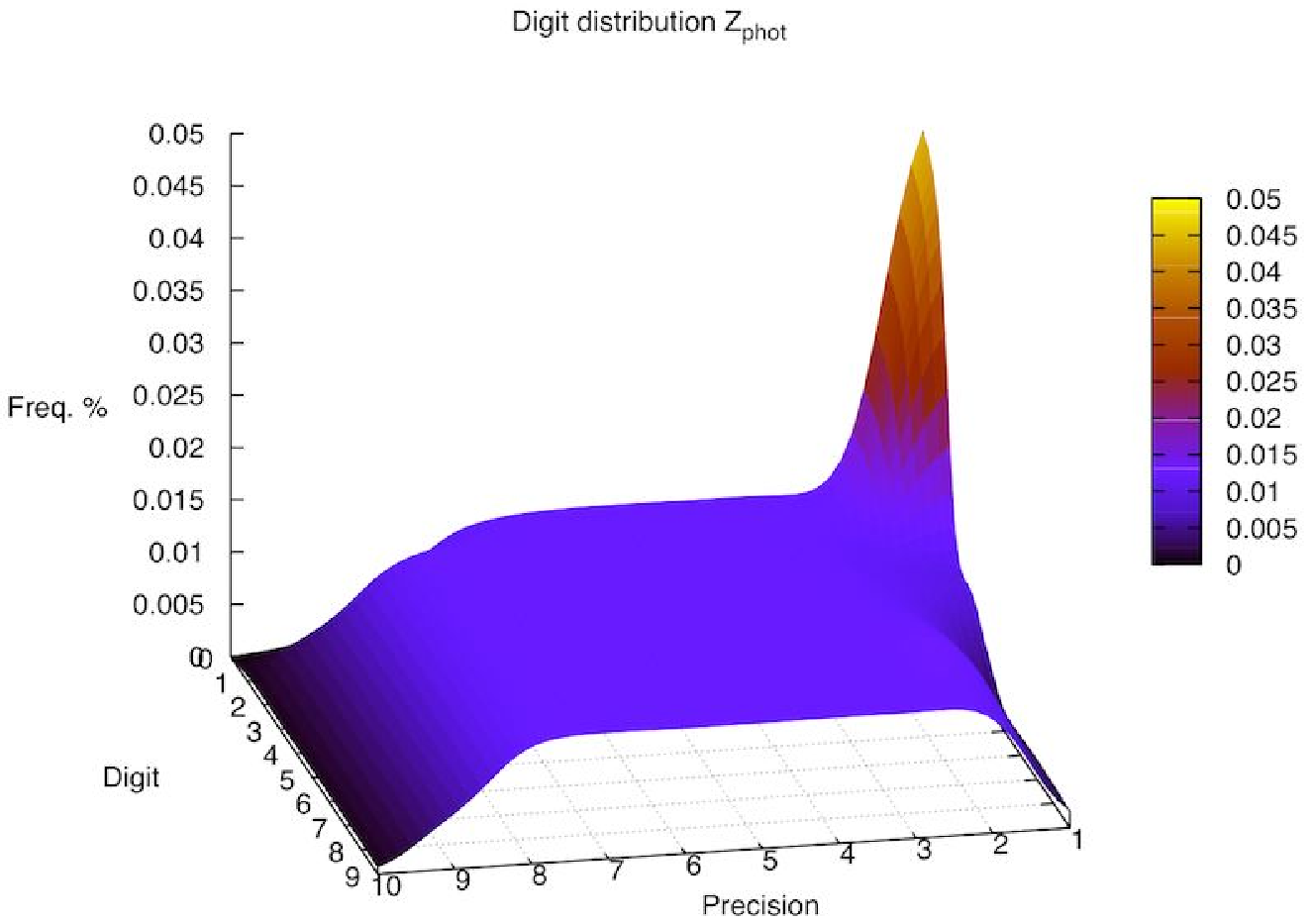}
    \caption{Digit distribution for $z_{spec}$ and $z_{phot}$; Top: Spectrometric digit distribution; Bottom: Photometric digit distribution. Note that digit distribution for $z_{spec}$ has three peaks, but $z_{phot}$ has only one. }
  \label{fig:3D}
  \end{center}
\end{figure}

With this tree it is a trivial task to build bins for data distribution.  Figure~\ref{fig:3D} depicts the frequency distribution for a given digit and precision. There are 100 data points that have been convolved with a Gaussian kernel to produce surface planes in order to assemble three-dimensional plots.

This helps to build a cluster-wise mapping of the data. Following the Figure~\ref{fig:3D} top panel we observe that for the first decimal digit most data observations are concentrated in the digits 0, 1, 2, and 3. Then the rest of decimal precision data is uniformly distributed, gradually going towards zero when the level of precision increases. There is the exception of two peaks, for precision equal to 8. This turns out to be useful because when comparing the $z_{spec}$ and $z_{phot}$ digit distribution we do not find the same peaks in $z_{phot}$.  This is very useful because now we can discriminate which observations are more reliable in  $z_{phot}$ through different characteristics of the data associated with the peaks. 

\clearpage

\section{Conclusions}

In this work the distance called the Baire distance is presented.  This distance has been very recently introduced into data analysis.  We show how this distance can be used to generate clusters in a way that is computationally inexpensive when compared with more traditional techniques. The distance is also an ultrametric and therefore furnishes an ultrametric too.   Hence the new Baire distance lends itself very well to the new hierarchical clustering method that we have introduced here.   

In the astronomy case clusters generated with the Baire distance can be useful when calibrating redshifts. In general, applying the Baire method to cases where digit precision is important  can be of relevance, specifically to  highlight data ``bins'' and some of their properties.

Note that when two numbers share 3 prefix digits, and base 10 is used, we have a Baire distance of $\db = 10^{-3}$. We may not need to define the actual (ultra)metric values. It may be, in fact, more convenient to work on the hierarchy, with its different levels.

In section~\ref{subsection:clustering-based-baire} we showed how we could derive that $82.8\%$ of values for $z_{spec}$ and $z_{phot}$ have at least two common prefix digits. This is a powerful result in practice when we recall that we can find very efficiently where these $82.8\%$ of the astronomical objects are. 

Using the Baire distance we showed in section~\ref{section:digit-distribution} that $z_{spec}$ and $z_{phot}$ signals can be stored in a tree like structure. This is advantageous when measuring the digit distribution for each signal.  When comparing these distributions, it can easily be seen where the differences arise.

The Baire distance has proved very useful in a number of cases, for instance 
in Murtagh et al.~\cite{Murtagh08} this distance is used in conjunction with random projection~\cite{vempala04} as the basis for clustering 
a large dataset of chemical compounds achieving results comparable to $k$-means but with better performance due to the lower computational complexity of the Baire-based clustering method.

Other application areas include text mining and semantic preservation~\cite{Pereira00}.  For more details refer to Contreras~\cite{pedrophd} where a number of examples are discussed.

\bibliographystyle{plain}
\bibliography{biblio-preprint}

\end{document}